\documentclass[%
reprint,
superscriptaddress,
 amsmath,amssymb,
 aps,
prb,
longbibliography,
]{revtex4-1}
\usepackage{textcomp}
\usepackage{graphicx}% Include figure files
\usepackage{dcolumn}% Align table columns on decimal point
\usepackage{bm}% bold math
\usepackage{color}
\usepackage{lineno}
\begin{document}

\preprint{APS/123-QED}

\title{
Direct observation of current-induced nonlinear spin torque in Pt-Py bilayers}

% Force line breaks with \\
%\thanks{A footnote to the article title}%

\author{Toshiyuki Kodama}
\email[Email address:]{tkodama@tohoku.ac.jp}
\affiliation{Institute for Excellence in Higher Education, Tohoku University, Sendai, 980-8576, Japan}

\author{Nobuaki Kikuchi}
\affiliation{Institute of Multidisciplinary Research for Advanced Materials, Tohoku University, Sendai 980-8577, Japan}

\author{Takahiro Chiba}
\affiliation{Frontier Research Institute for Interdisciplinary Sciences, Tohoku University, Sendai 980-8578, Japan}

\author{Satoshi Okamoto}
\affiliation{Institute of Multidisciplinary Research for Advanced Materials, Tohoku University, Sendai 980-8577, Japan}
\affiliation{Center for Science and Innovation in Spintronics, Tohoku University, Sendai 980-8577, Japan}

\author{Seigo Ohno}
\affiliation{Department of Physics, Graduate School of Science, Tohoku University, Sendai 980-8578, Japan}

\author{Satoshi Tomita}
\email[Email address:]{tomita@tohoku.ac.jp}
\affiliation{Institute for Excellence in Higher Education, Tohoku University, Sendai, 980-8576, Japan}
\affiliation{Department of Physics, Graduate School of Science, Tohoku University, Sendai 980-8578, Japan}

%\date{\today}% It is always \today, today,
             %  but any date may be explicitly specified

\begin{abstract}
We experimentally observe 
nonlinear spin torque 
in metallic bilayers of platinum and permalloy 
by means of spin-torque ferromagnetic-resonance (ST-FMR) 
under massive dc current injection.
The observed nonlinear spin torque 
exerted to permalloy magnetization 
is attributed primarily to 
nonlinear spin polarization.
Additional origin of the nonlinear spin torque 
is magnon generation (annihilation) 
followed by shrinkage (expansion) of effective magnetization,
which is reveled by ST-FMR and unidirectional spin Hall magnetoresistance measurements.
The present study 
paves a way to 
spin-Hall effect based nonlinear spintronic devices 
as well as time-varying nonlinear magnetic metamaterials 
with tailor-made permeability.
\end{abstract}

%\keywords{Suggested keywords}%Use showkeys class option if the keyword
                              %display desired
\maketitle

\section{Introduction}
Nonlinear phenomena 
are common and intriguing topics 
in condensed matter physics \cite{Robert2008}.
Among the fascinating achievements 
in spintronics and magnonics are 
nonlinear spin torque oscillator \cite{Yamaguchi2019-xt, Iwakiri2020-ac}, 
nonlinear spin-wave interference 
toward memory or neural network systems \cite{Adhikari2020-aa, Hula2022-jh}, 
and nonlinear magnon polariton 
for quantum information technologies \cite{Lee2023-ar}.
Recently, 
several theoretical studies 
predicted nonlinear spin polarization in  
non-centrosymmetric system \cite{Hamamoto2017},
$\mathcal{P}\mathcal{T}$-symmetric collinear magnets \cite{Hayami2022-kt}, and 
time-reversal centrosymmetric materials \cite{Xiao2022PRL, Xiao2023PRL}.
The nonlinear spin polarization is of great interest 
because it probes novel band geometric quantities 
and offers new tools to characterize and control material properties.
However, 
lacking is 
experimental studies of 
the nonlinear spin polarization.
Therefore, in this work, 
we investigate experimentally
the nonlinear spin polarization 
using spin-torque ferromagnetic-resonance (ST-FMR).

When electric current flows in a bilayer system 
consisting of heavy-metal, for example platinum (Pt), 
and ferromagnetic-metal, for example, permalloy (Py), 
the spin-Hall effect 
due to strong spin-orbit interaction in the Pt layer 
gives rise to spin polarization.
The spin polarization causes 
spin current injected to the Py layer, 
bringing about spin torque 
exerted to precessing Py magnetization 
on resonance under magnetic fields; 
this is referred to as ST-FMR \cite{Liu2011-vd}. 
In this paper, 
we carry out ST-FMR measurements
under large dc current injection 
up to 20 mA  ($\sim 6.5\times10^{11}$ A/m$^{2}$) 
to directly observe 
current-induced nonlinear spin torque in Pt-Py bilayers.
An undoped silicon (Si) substrate with excellent thermal conductivity 
enables us to inject such a large current
without sample degradation due to the Joule heating.
The ST-FMR signals demonstrate that 
the massive dc current 
affects the resonance field and Gilbert damping parameter nonlinearly. 
The nonlinear changes 
are traced back to 
the nonlinear spin torque 
caused by the nonlinear spin polarization.
Furthermore, 
ST-FMR study reveals that 
the nonlinear spin torque 
is attributed also to
magnon generation (annihilation) 
followed by effective magnetization shrinkage (expansion), 
which is confirmed by the observation of 
unidirectional spin Hall magnetoresistance (USMR) 
\cite{Avci2015}.

Eventually 
we evaluate the origins of the nonlinear spin torque,
i.e., the nonlinear spin polarization and magnon generation/annihilation,
by introducing indices of nonlinearity, $\eta$ and $\xi$, 
obtained from ST-FMR and USMR measurements.
The experimentally evaluated $\eta$ is larger than $\xi$, 
indicating that 
the nonlinear spin polarization is dominant 
rather than the magnon generation/annihilation 
in the nonlinear spin torque.
The $\eta$ and $\xi$ correspond respectively to  
the 2nd- and 3rd-order nonlinear susceptibilities, 
$\chi^{(2)}$ and $\chi^{(3)}$,
in nonlinear photonics \cite{Robert2008}.
In analogy between photonics and electronics,
$\eta$ and $\xi$ can be used 
to evaluate spintronic nonlinearity,
elucidate the origins of nonlinear phenomena, 
and realize nonlinear spintronic effects, 
for example, second harmonic generation and rectification.
Furthermore, 
the nonlinear spin torque 
leads to time-varying nonlinear magnetic metamaterials 
for 6th-generation mobile communication light sources 
of millimeter waves and THz light \cite{Kodama2023}.

\section{Experimental setup}
\begin{figure}[tb!]
\includegraphics[width=8.6truecm,clip]{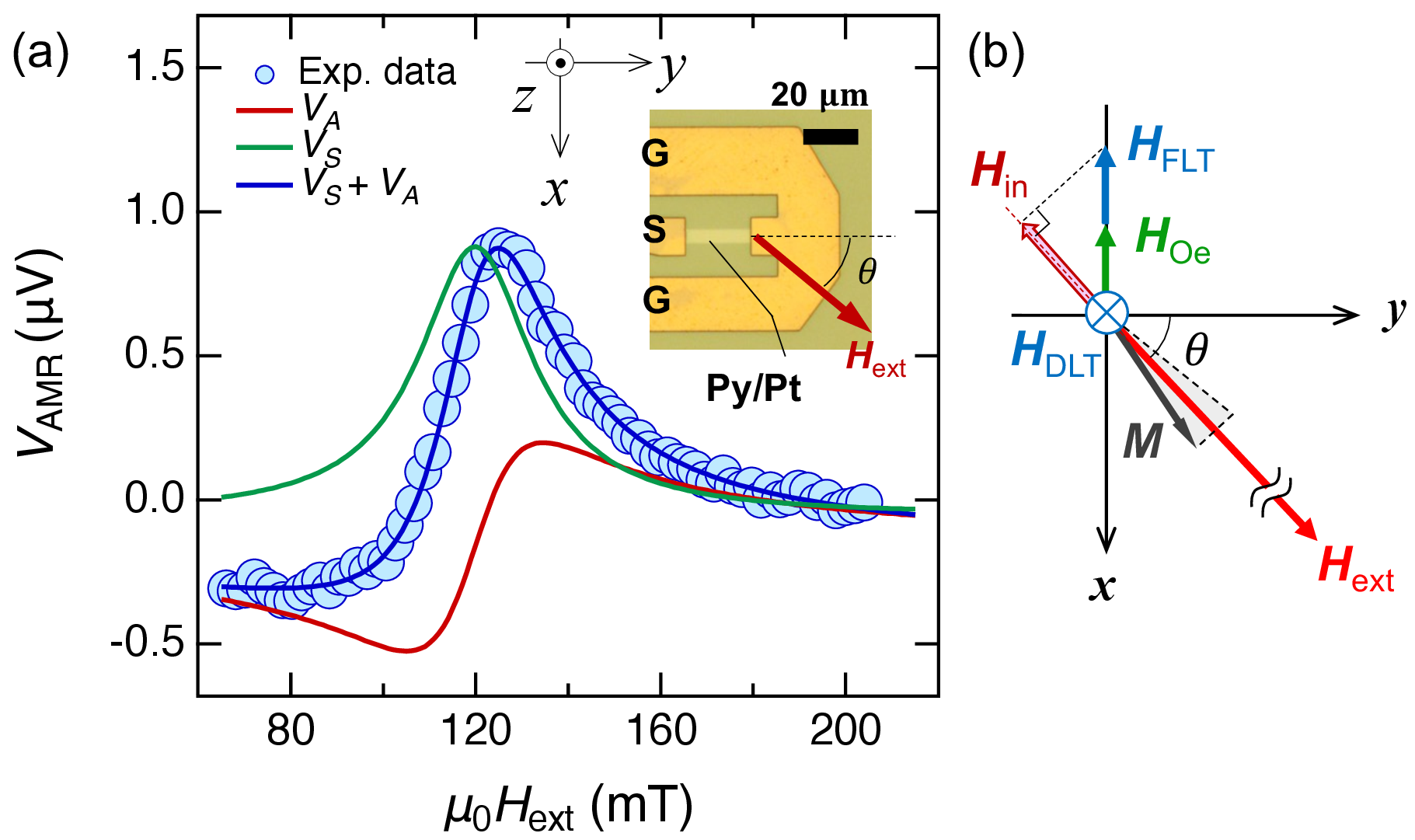}
\caption{
(a) Measured ST-FMR signal (blue circles), 
as a function of external dc magnetic field, $\mu_{0} H_{\rm ext}$, 
with $I_{\rm ac}$ at 9 GHz. 
Green and red solid lines 
represent fitting curves
with symmetric ($V_{\rm S}$) 
and anti-symmetric coefficients ($V_{\rm A}$), respectively.
Blue solid line 
corresponds to the sum of $V_{\rm S}$ and $V_{\rm A}$.
Inset: optical microscopic image 
of Py-Pt bilayer strip with Au electrodes.
(b) Schematic of 
Py magnetization ($M$), 
$H_{\rm ext}$, 
and Oersted field ($H_{\rm Oe}$). 
Effective field 
$H_{\rm FLT}$ and $H_{\rm DLT}$
correspond to field-like torque and damping-like torque, respectively.
In-plane effective field $H_{\rm in}$ is
a component of $H_{\rm FLT}$ and $H_{\rm Oe}$
parallel to $H_{\rm ext}$.}
\label{stfmr}
\end{figure}

We study metallic bilayers 
composed of 5 nm thick Pt top layer 
and 2 nm thick Py bottom layer.
The Pt-Py bilayer is deposited after 
a 3 nm thick tantalum buffer layer
on an undoped Si substrate 
having electrical resistivity at least 1 k$\Omega\cdot$cm (Crystal Base, Inc.) \cite{Kodama2023}. 
An inset of Fig. \ref{stfmr}(a) shows 
an optical microscopic image of the specimen 
consisting of lithographically-prepared Pt-Py strip 
attached to gold electrodes.
The width of the strip is 5 $\mu$m and the length is 24 $\mu$m.
In ST-FMR measurements, 
an in-plane external dc magnetic field $H_{\rm ext}$ 
is applied with a relative angle $\theta = 45^\circ$ to the $y$-axis 
as shown in Figs. \ref{stfmr}(a) and \ref{stfmr}(b). 
An ac current $I_{\rm ac}$ with microwave frequencies
is applied 
between the signal (S) and ground (G) lines 
by a signal generator.
The $I_{\rm ac}$ in the Pt layer 
generates an oscillating Oersted magnetic field, 
which primarily drives ST-FMR of the Py magnetization $M$. 
Additionally, 
the spin-Hall effect in the Pt layer 
gives rise to ac spin current, 
which is injected into the Py layer.
The spin angular momentum 
is transferred to the in-plane Py magnetization, 
exerting a field-like torque (FLT) 
that secondary drives ST-FMR 
and a damping-like torque (DLT) 
that enhances or reduces magnetic relaxation 
\cite{chiba2014prappl, chiba2015jap, schreier2015prb}.
Mixing of $I_{\rm ac}$ and 
oscillating anisotropic magnetoresistance (AMR) in Py 
gives rise to a time-independent longitudinal dc voltage $V_{\rm AMR}$.
We measure $V_{\rm AMR}$ 
as a function of $\mu_{0} H_{\rm ext}$ 
using a bias tee 
to obtain ST-FMR signals.
All measurements are carried out at room temperature.

\section{Results}
\subsection{Spin-torque ferromagnetic resonance measurement}
Figure \ref{stfmr}(a) 
shows a typical ST-FMR signal 
probed by $V_{\rm AMR}$ 
with $I_{\rm ac}$ at 9 GHz.
The blue circles 
correspond to measured $V_{\rm AMR}$. 
The $V_{\rm AMR}$ in a thin film
is expressed as 
$V_{\rm AMR} = V_{\rm S} + V_{\rm A}$, 
where $V_{\rm S}$ and $V_{\rm A}$ are symmetric and anti-symmetric components, 
respectively \cite{Liu2011-vd}.
Both $V_{\rm S}$ and $V_{\rm A}$ 
are described using
$\mu_{0} H_{\rm ext}$, 
the resonance field ($\mu_{0} H_{\rm FMR}$),
and the half width at half maximum ($\mu_{0} \Delta_{\rm FMR}$) 
of the FMR signal.
The fitting in Fig. \ref{stfmr}(a)  
gives $\mu_{0} H_{\rm FMR}$ = 119.8 mT 
and $\mu_{0} \Delta_{\rm{FMR}}$ = 14.8 mT. 
The sum of 
$V_{\rm S}$ (green line) and $V_{\rm A}$ (red line) 
after the fitting
is represented by the blue line, 
which reproduces well the measured $V_{\rm AMR}$.

Together with $I_{\rm ac}$,
a dc current $I_{\rm dc}$ 
is applied to the bilayer 
to modify the FMR condition \cite{Kasai2014-ia, Nan2015-yg}.
The $I_{\rm dc} > 0$ ($I_{\rm dc} < 0$) 
corresponds to 
the current in the $+y$ ($-y$) direction.
The $I_{\rm dc}$ causes 
a time-independent dc Oersted field $\bm H_{\rm Oe}$ along $\pm x$ axis 
as shown in Fig. \ref{stfmr}(b). 
Additionally, 
$I_{\rm dc}$ 
generates a time-independent FLT and DLT on $\bm M$.
As in Fig. \ref{stfmr}(b), 
FLT and DLT
are regarded as effective fields 
$\bm H_{\rm FLT} \propto {\delta}{\bm s}$ and $\bm H_{\rm DLT} \propto {\bm m} \times {\delta}{\bm s}$, 
respectively.
The ${\delta}{\bm s}$  is spin polarization and
${\bm m}$ is a unit vector of magnetization
\cite{Fan2013-tq, Karube2020-df}.
The $\bm H_{\rm FLT}$ and $\bm H_{\rm DLT}$ 
affect the FMR condition, 
resulting in a shift of $\mu_{0} H_{\rm FMR}$ 
and a change in $\mu_{0} \Delta_{\rm FMR}$.

\begin{figure}[tb!]
\includegraphics[width=8.0truecm,clip]{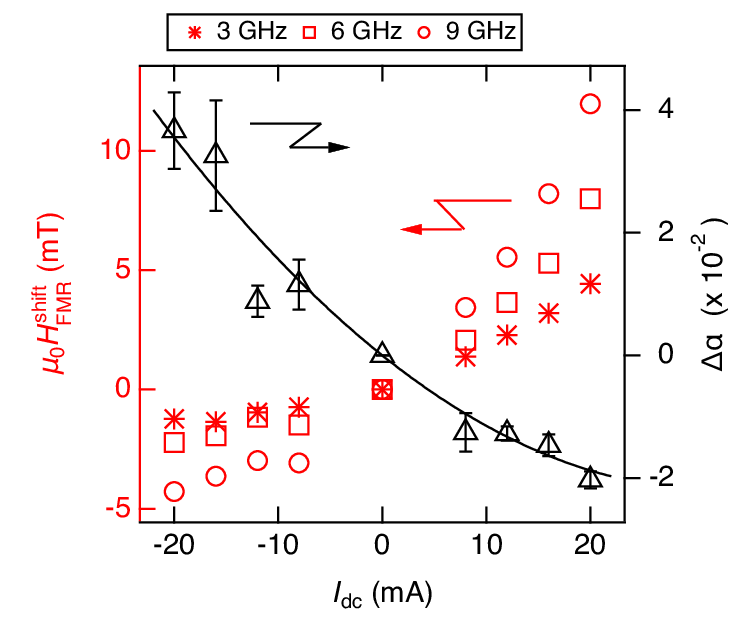}
\caption{
$I_{\rm dc}$ versus resonance field shift $\mu_{0} H^{\rm shift}_{\rm FMR}$ (left axis) 
with $I_{\rm ac}$ at 3 GHz (red asterisks), 6 GHz (red squares), and 9 GHz (red circles). 
Gilbert damping parameter variation $\Delta \alpha$  (right axis, black triangles) 
reproduced from our previous report \cite{Kodama2023} 
is also plotted as a function of $I_{\rm dc}$.
The solid line corresponds to the fitting curve.}
\label{result}
\end{figure}

To study the shift of $\mu_{0} H_{\rm FMR}$ 
and change in $\mu_{0} \Delta_{\rm FMR}$ 
by $\bm H_{\rm FLT}$ and $\bm H_{\rm DLT}$, 
the ST-FMR signals 
with $I_{\rm dc}$ between $-$20 and $+$20 mA 
are measured at various $I_{\rm ac}$ frequency ($f_{\rm ac}$) 
from 3 to 9 GHz. 
Thanks to the undoped Si substrate  
with a better thermal conductivity of 150 W/mK \cite{Slack1964-pl} 
compared to quartz (1.4 W/mK) \cite{Zhu2018-ib} 
and magnesium oxide (56 W/mK) \cite{Stackhouse2010-ph} substrates,
a large $I_{\rm dc}$ up to $\pm$20 mA 
can be applied 
(See Supplemental Material SM1 \cite{SM}).
After the fitting of the ST-FMR signals, 
we evaluate $\mu_{0} H_{\rm FMR}$ and $\mu_{0}\Delta_{\rm FMR}$ 
at a specific $I_{\rm dc}$ value.
The resonance field shift $\mu_{0} H^{\rm shift}_{\rm FMR}$ 
by $I_{\rm dc}$ injection
is derived from 
$\mu_{0} H^{\rm shift}_{\rm FMR}(I_{\rm dc}) 
= \mu_{0} H_{\rm FMR}(I_{\rm dc}) - \mu_{0} H_{\rm FMR}(0)$, 
where $\mu_{0} H_{\rm FMR}(I_{\rm dc})$ 
corresponds to $\mu_{0} H_{\rm FMR}$ 
at non-zero $I_{\rm dc}$ 
and $\mu_{0} H_{\rm FMR}(0)$ 
corresponds to $\mu_{0} H_{\rm FMR}$ 
at zero $I_{\rm dc}$.
Moreover, 
$f_{\rm ac}$-dependence 
of $\mu_{0}\Delta_{\rm FMR}$ 
gives Gilbert damping parameter $\alpha$ 
at a specific $I_{\rm dc}$ value.

Figure \ref{result} shows 
$I_{\rm dc}$ versus $\mu_{0} H^{\rm shift}_{\rm FMR}$
at $f_{\rm ac}$ = 3 GHz (red asterisks), 6 GHz (red squares), and 9 GHz (red circles)  
as indicated from the left vertical axis.
In addition, 
the variation in $\alpha$ by $I_{\rm dc}$ injection, 
$\Delta \alpha (I_{\rm dc}) = \alpha (I_{\rm dc}) - \alpha(0)$, 
reproduced from our previous report \cite{Kodama2023}, 
is plotted as black triangles indicated from the right vertical axis.
The $\mu_{0} H^{\rm shift}_{\rm FMR}$ 
and $\Delta \alpha$
are odd functions of $I_{\rm dc}$,
because $\delta{\bm s}$ is odd function of the dc current.
Figure \ref{result} highlights two striking features:
i) $\mu_{0} H^{\rm shift}_{\rm FMR}$ and $\Delta \alpha$
are dependent nonlinearly on $I_{\rm dc}$, 
and 
ii) a higher $f_{\rm ac}$ 
results in a larger $|\mu_{0} H^{\rm shift}_{\rm FMR}|$ at the same $I_{\rm dc}$.
Note that these features are observed
in another specimen with a longer Pt-Py strip of 45 $\mu$m length
(see Supplemental Material SM2 \cite{SM}).

\subsection{Evaluation of variation in effective magnetization}

\begin{figure}[tb!]
\includegraphics[width=8truecm,clip]{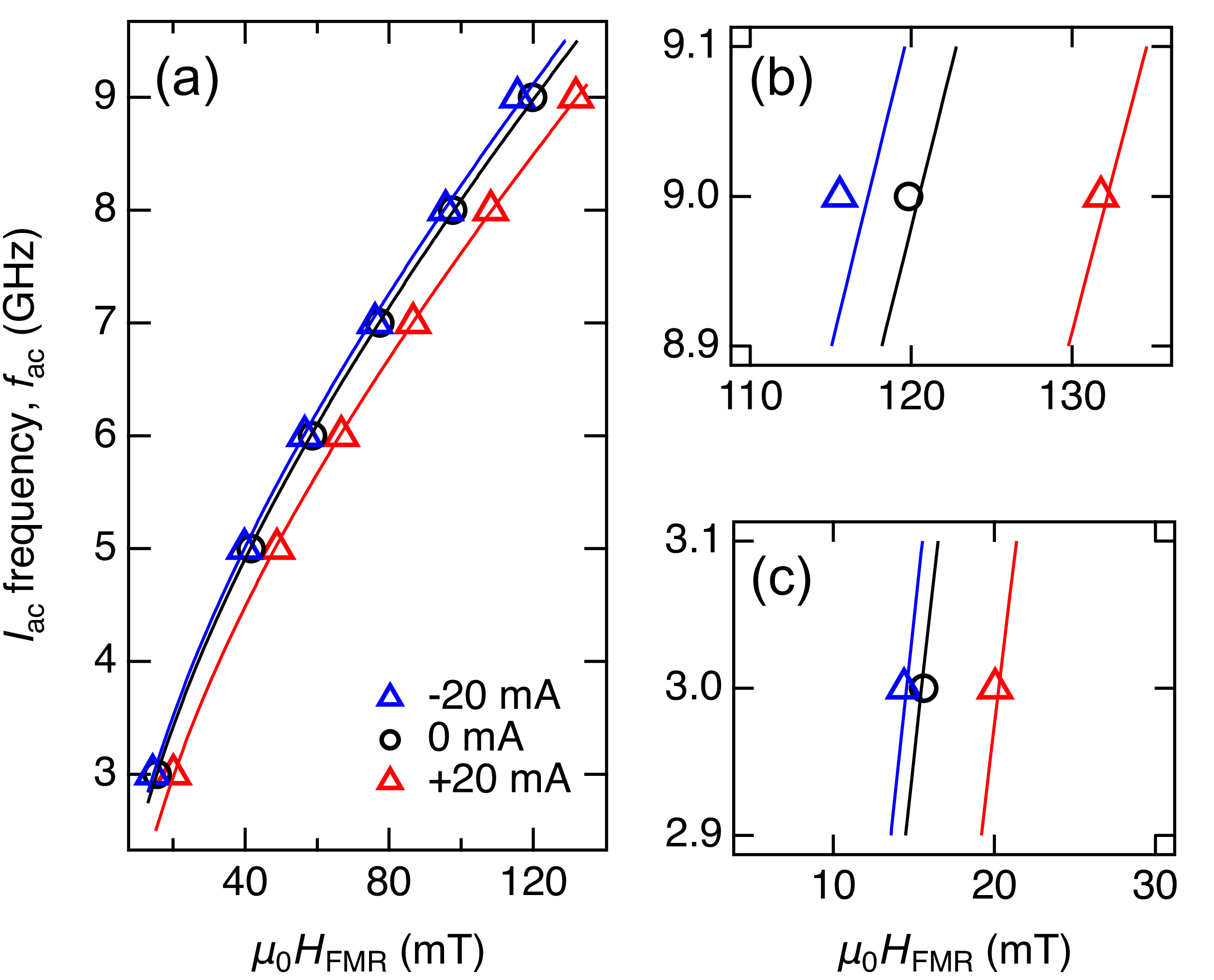}
\caption{
(a) $I_{\rm ac}$ frequency ($f_{\rm ac}$) versus 
$\mu_{0} H_{\rm FMR}$ 
at $I_{\rm dc}$ = $-$20 mA (blue triangles), 0 mA (black circles), and +20 mA (red triangles). 
Solid lines: fitting curves using the Kittel equation [Eq. \eqref{kittel}]. 
(b) and (c) are an enlarged figure of (a).}
\label{kittelplot}
\end{figure}

The $\mu_{0} H_{\rm FMR}$
with various $I_{\rm dc}$ of $-$20, 0, and 20 mA
are plotted as a function of $f_{\rm ac}$ from 3 to 9 GHz 
in Fig. \ref{kittelplot}.
Blue triangles, black circles, and red triangles 
correspond respectively to $\mu_{0} H_{\rm FMR}$ 
by $I_{\rm dc}$ = $-$20, 0, and 20 mA.
Figures \ref{kittelplot}(b) and \ref{kittelplot}(c) 
show enlarged plots of Fig. \ref{kittelplot}(a), 
in which the horizontal axis variations 
are identical to be 25 mT for direct comparison. 
At $f_{\rm ac}$ = 3 GHz 
as in Fig. \ref{kittelplot}(c), 
the resonance field shifts upward by 5.7 mT
when $I_{\rm dc}$ increases from $-$20 mA to 20 mA. 
At a higher $f_{\rm ac}$ of 9 GHz 
as in Fig. \ref{kittelplot}(b), 
the shift amount is larger to be 16.2 mT.

As shown in Fig. \ref{stfmr}(b), 
a component of $H_{\rm FLT}$ and $H_{\rm Oe}$ 
parallel to the $H_{\rm ext}$ 
corresponds to the in-plane effective field $H_{\rm in}$.
The $H_{\rm in}$ expressed as
$\mu_{0} H_{\rm in} = (\mu_{0} H_{\rm FLT} + \mu_{0} H_{\rm Oe}) \times \sin \theta$, 
where $\theta = \pi/4$ in the present ST-FMR study, 
is small, 
but affects the FMR condition.
(see Supplemental Material SM3 \cite{SM}).
The Kittel equation 
for FMR is described as
\begin{eqnarray}
\label{kittel}
2 \pi f_{\rm ac} 
= \gamma \sqrt{\mu_{0} H_{\rm FMR} (I_{\rm dc} = 0) + \mu_{0} H_{\rm in}} \nonumber \\
\times \sqrt{\mu_{0} H_{\rm FMR} (I_{\rm dc} = 0) + \mu_{0} H_{\rm in} + \mu_{0} M_{\rm eff}}, 
\end{eqnarray}
where $\gamma$ 
is gyromagnetic ratio, 
$\mu_{0} H_{\rm FMR} (I_{\rm dc} = 0)$
is resonance field without $I_{\rm dc}$ injection, and
$\mu_{0} M_{\rm eff}$ is effective magnetization.
When $I_{\rm dc}$ = 0 mA,
the black circles in Fig. \ref{kittelplot} 
are fitted by Eq. \eqref{kittel} 
with $\mu_{0} H_{\rm in}  = 0$.
The fitting gives 
$\mu_{0} M_{\rm eff}(I_{\rm dc}=0) = \mu_{0} M_{\rm{s}} =$ 658 mT, 
where $\mu_{0} M_{\rm{s}}$ is the saturation magnetization.
The $ \mu_{0} M_{\rm{s}}$ of 658 mT 
is smaller than a typical value of Py saturation magnetization, 
probably due to magnetic dead layers 
at the Pt-Py interface \cite{Kodama2023, Hirayama2017-kz}.

\begin{figure}[tb!]
\includegraphics[width=7truecm,clip]{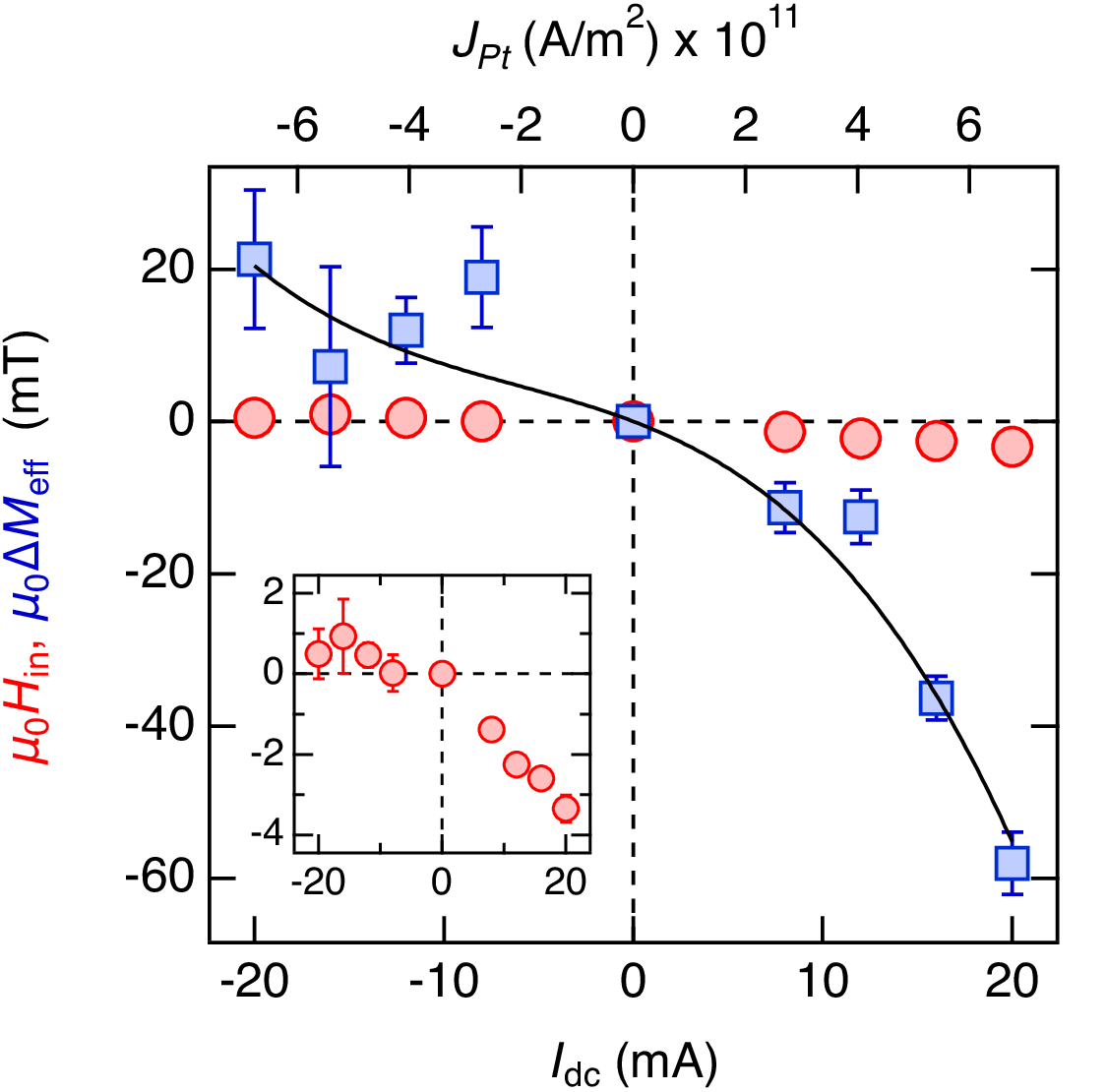}
\caption{
In-plane effective field $\mu_{0} H_{\rm in}$ (red circles)
and effective magnetization variation $\mu_{0}\Delta M_{\rm eff}$ (blue squares) 
are plotted as a function of $I_{\rm dc}$. 
Inset: enlarged view for $\mu_{0} H_{\rm in}$.
The current density in Pt layer converted from $I_{\rm dc}$ 
is indicated from upper horizontal axis.
Solid line corresponds to a fitting curve. 
Error bars are the standard deviation of the fitting.}
\label{anis}
\end{figure}

Equation \eqref{kittel} 
is fitted to $\mu_{0} H_{\rm FMR}$ 
at $I_{\rm dc} =$ $\pm$ 8, $\pm$ 12, $\pm$ 16, and $\pm$ 20 mA 
to evaluate $\mu_{0} M_{\rm eff}$ and $\mu_{0} H_{\rm in}$.
The fitting curves with $I_{\rm dc}$ = $-$20 and +20 mA
are drawn by blue and red solid lines in Fig. \ref{kittelplot}, respectively.
In Fig. \ref{anis}, 
evaluated $\mu_{0} H_{\rm in}$ (red circles)
and $\mu_{0} \Delta M_{\rm eff} = \mu_{0} M_{\rm eff} - \mu_{0} M_{\rm s}$ (blue square) 
are plotted as a function of $I_{\rm dc}$.
The upper horizontal axis indicates the $J_{\rm Pt}$,
electrical current density in the Pt layer 
(see Supplemental Material SM4 \cite{SM}).
The inset shows an enlarged view
of $\mu_{0} H_{\rm in}$ versus $I_{\rm dc}$.
The $\mu_{0} H_{\rm in}$ is very small, 
slightly decreases with increasing $I_{\rm dc}$,
and reaches at $-3.3$ mT when $I_{\rm dc}$ is 20 mA.

Contrastingly, 
$\mu_{0}\Delta M_{\rm eff}$ 
is affected significantly by $I_{\rm dc}$.
The maximum value of $\mu_{0}\Delta M_{\rm eff}$ 
at $I_{\rm dc} =$ 20 mA
is $-60$ mT, 
which includes $\mu_{0} H_{\rm DLT}$ 
as shown in Fig. \ref{stfmr}(b).
However, 
$\mu_{0}\Delta M_{\rm eff}$ = $-60$ mT
is nonetheless larger than 
the $\mu_{0} H_{\rm DLT}$ evaluated from previous reports,
for example,
$-2.4$ mT 
at approximately $I_{\rm dc} =$ 20 mA 
($ \sim 6.5 \times 10^{11} \rm A/m^{2})$
in Refs. \cite{Karube2020-df}.
Note here that 
$\mu_{0} H_{\rm FMR}$ 
with non-zero $I_{\rm dc}$
in Fig. \ref{kittelplot} 
cannot be reproduced 
using Eq. \eqref{kittel} without $\mu_{0}\Delta M_{\rm eff}$.
This is clearly indicated in Eq. \eqref{kittel}, i.e., 
$\mu_{0} H_{\rm in}$ shifts the curves 
whereas $\mu_{0}\Delta M_{\rm eff}$ changes the gradient of the curves. 
The large $\mu_{0}\Delta M_{\rm eff}$
is indispensable
in explaining a larger resonance field shift
at a higher $f_{\rm ac}$.
The negative $\mu_{0}\Delta M_{\rm eff}$ 
corresponds to 
shrinkage of $\mu_{0} M_{\rm eff}$,
whereas 
the positive $\mu_{0}\Delta M_{\rm eff}$ 
corresponds to expansion.
Therefore, 
we consider 
shrinkage/expansion of effective magnetization 
by nonlinear magnon generation/annihilation.

\begin{figure}[tb!]
\includegraphics[width=8.6truecm,clip]{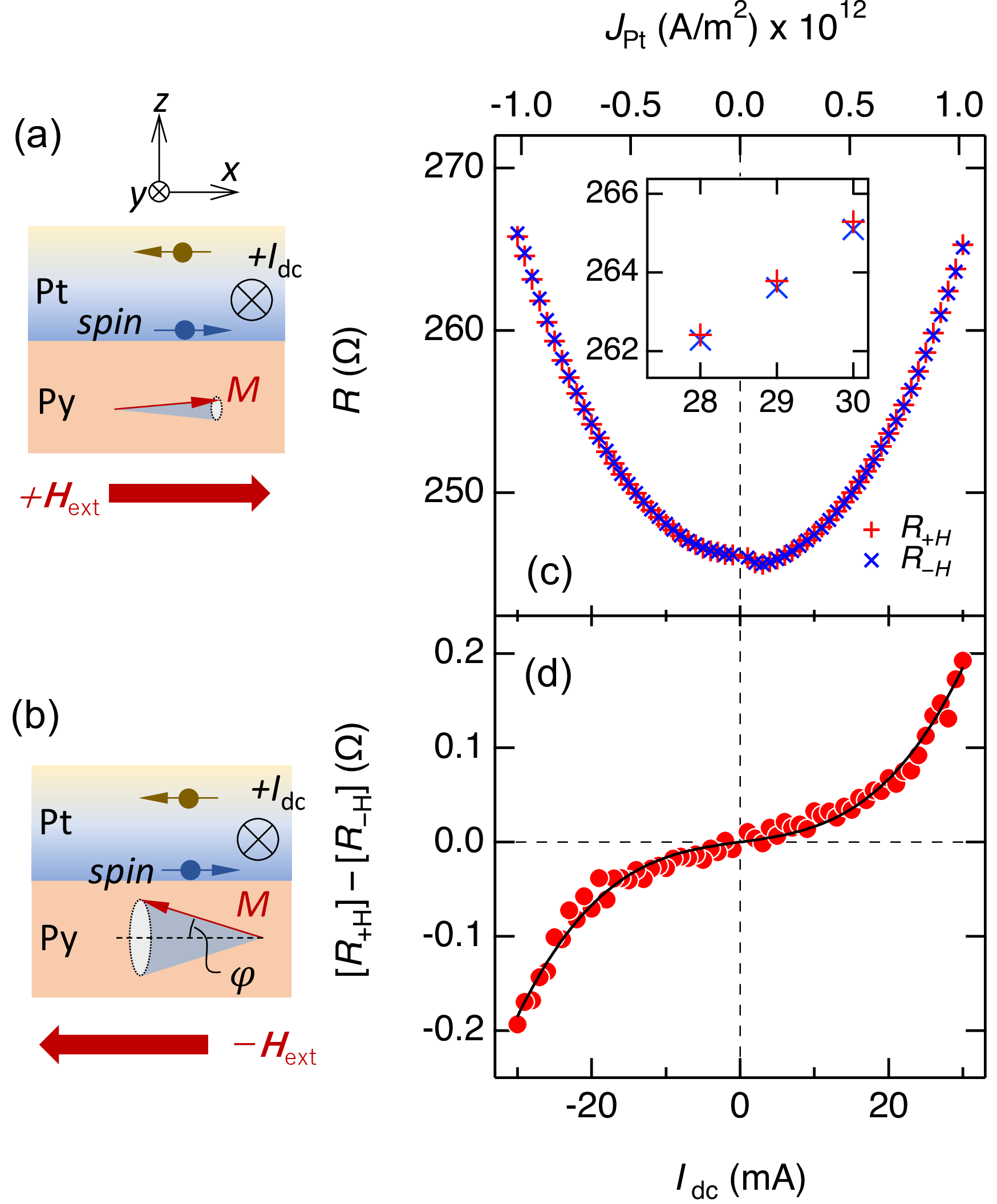}
\caption{
Schematics of specimen cross-section for  
(a) $R_{\rm +H}$ under $\mu_{0} H_{\rm ext}$ = $+$100 mT and 
(b) $R_{\rm -H}$ under $\mu_{0} H_{\rm ext}$ = $-$100 mT.
(c) $I_{\rm dc}$ versus electric resistance 
$R_{\rm +H}$ and $R_{\rm -H}$ .
The current density in Pt layer converted from $I_{\rm dc}$ 
is indicated from upper horizontal axis.
Inset: Enlarged view around 30 mA.
(d) Difference between the electric resistance 
under $\mu_{0} H_{\rm ext}$ = $+$100 mT and $-$100 mT, 
$[R_{\rm +H}]-[R_{\rm -H}]$ is plotted as a function of $I_{\rm dc}$.
The solid line corresponds to the fitting curve.}
\label{usmr}
\end{figure}

\begin{figure}[tb!]
\includegraphics[width=6.5truecm,clip]{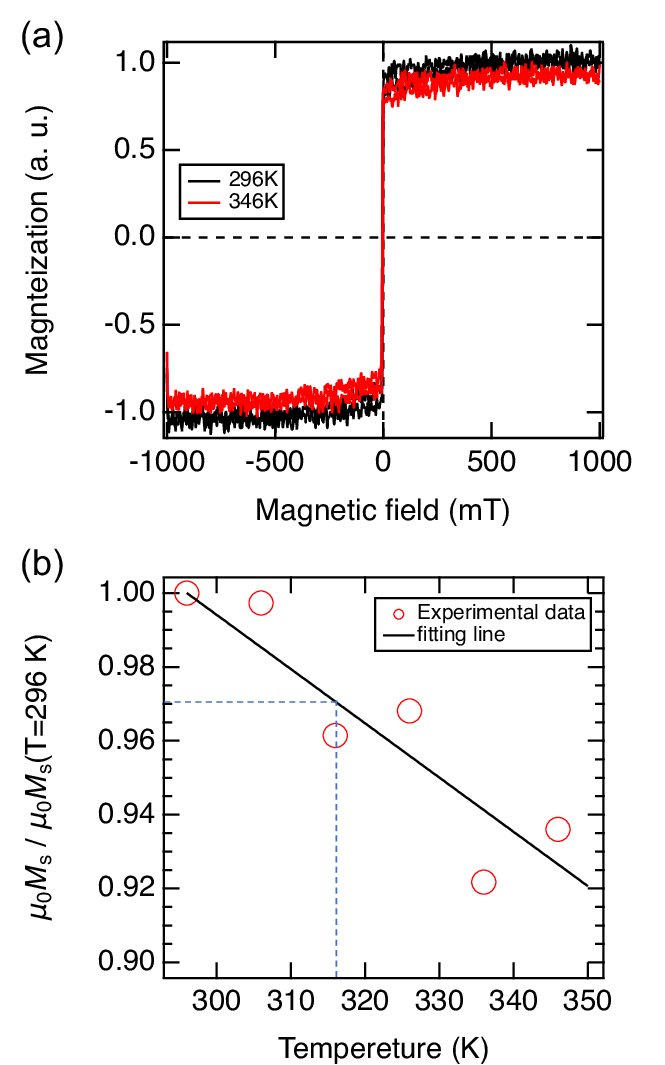}
\caption{
(a) Magnetization curves of Pt-Py bilayer 
at 296 (black) and 346 K (red) measured using VSM. 
(b) Normalized saturation magnetization $\mu_{0}M_{\rm S}$ (red circles) 
as a function of temperature. 
The black line shows a fitting result using a linear function.
The blue dotted line indicates that at 316 K, 
which is 20 K higher than room temperature,
normalized $\mu_{0}M_{\rm S}$ is 0.97.
}
\label{mh}
\end{figure}

\subsection{Unidirectional spin-Hall magnetoresistance measurement}
The nonlinear magnon generation/annihilation 
due to massive spin current injection
likely brings about USMR \cite{Borisenko2018APL, Kim2019APEX}. 
We thus conduct USMR measurements 
with the large $I_{\rm dc}$ injection 
using the same specimen. 
Figures \ref{usmr}(a) and \ref{usmr}(b) 
show schematics of the sample cross section 
in the $x$-$z$ plane viewed to the $+y$ direction.
The $\theta$ is $+ 90^{\circ}$ or $- 90^{\circ}$ 
in the USMR measurements, 
i.e.,the $\mu_{0} H_{\rm ext}$ = $+ 100$ mT 
is applied to the sample in the $+ x$ direction (Fig. \ref{usmr}(a)), 
while $\mu_{0} H_{\rm ext}$ = $- 100$ mT in the $- x$ direction (Fig. \ref{usmr}(b)).
The dc current $I_{\rm dc}$
between $-$30 mA and $+$30 mA 
is chopped to be 0.2 ms width pulses.
The longitudinal dc voltage $V_{\rm dc}$ 
are measured 100 times 
to obtain an averaged value of $V_{\rm dc}$.

The measured $V_{\rm dc} - I_{\rm dc}$ curve 
is converted to 
the dc electric resistance $R$ - $I_{\rm dc}$ curve  
as shown in Fig. \ref{usmr}(c).
Red crosses ($+$) 
correspond to 
$R$ under $\mu_{0} H_{\rm ext}$ = $+100$ mT, 
referred to as $R_{+ \rm H}$, 
while blue crosses ($\times$) corresponds to 
$R$ under $\mu_{0} H_{\rm ext}$ = $-100$ mT, 
referred to as $R_{- \rm H}$.
Because the current is applied beyond the ohmic region,
Fig. \ref{usmr}(c) shows 
a parabolic increase in $R_{+ \rm H}$ and $R_{- \rm H}$ 
due to the Joule heating \cite{Chiang2019-ju}.
Note here that 
a very similar increase in $V_{\rm dc}$  
is confirmed 
in a measurement using un-chopped continuous $I_{\rm dc}$ 
as same as in the ST-FMR study
(see Supplemental Material SM1 \cite{SM}).
Figure \ref{usmr}(c) shows that 
$R_{\rm \pm H}$ increases 
from 245 $\Omega$ at $| I_{\rm dc} | \sim$ 0 mA 
to 255 $\Omega$ at $| I_{\rm dc} | \sim$ 20 mA.
Given that 
a temperature coefficient of Pt resistance 
is 0.002 ${\rm K}^{-1}$, 
the increase in $R_{\rm \pm H}$ 
from 245 to 255 $\Omega$ 
corresponds to the sample temperature elevation 
of approximately 20 K \cite{Belser1959-pc}.
This evaluation clearly indicates that 
the Joule heating component is small and not dominant 
in the present ST-FMR experiments 
with $| I_{\rm dc} |$ up to 20 mA.

As highlighted in the inset of Fig. \ref{usmr}(c), 
an enlarged view at approximately $I_{\rm dc} =$ 30 mA,
$R_{\rm +H}$ (red crosses) 
is slightly larger than $R_{\rm -H}$ (blue crosses).
The difference between $R_{\rm +H}$ and $R_{\rm -H}$ 
is plotted as a function of $I_{\rm dc}$ in Fig. \ref{usmr}(d).
Note that 
the small Joule heating contribution, 
which is independent of the Py magnetization reversal,
is already removed in the $[R_{\rm +H}]-[R_{\rm -H}]$ plot.
As $I_{\rm dc}$ increases from 0 mA to 30 mA, 
$[R_{\rm +H}]-[R_{\rm -H}]$ 
increases slowly 
and then rapidly at above 20 mA.
More strikingly, 
$[R_{\rm +H}]-[R_{\rm -H}]$ 
is odd under the $I_{\rm dc}$ direction reversal;
this is the hallmark of USMR.

The USMR is caused by the electron scattering by magnons.
When the spins are injected into the Py layer, 
parallel spin injection to the Py magnetization
annihilates the magnons 
as in Fig. \ref{usmr}(a) 
whereas anti-parallel spin injection generates the magnons 
as in Fig. \ref{usmr}(b).
The magnon generation/annihilation 
influences the electron-magnon scattering, 
resulting in a resistance change of the Py layer as USMR.
The excited magnon number is increased nonlinearly 
when the inherent damping of Py is compensated 
by the anti-damping DLT \cite{Borisenko2018APL}.
This is consistent with 
nonlinear decrease in $\Delta \alpha$ observed in Fig. \ref{result}.
The magnon excitation depending on the current is expressed as 
$a_{\rm USMR} I_{\rm dc} + c_{\rm USMR} (I_{\rm dc})^{3}$,
where $a_{\rm USMR} $ and  $c_{\rm USMR} $ 
are linear and 3rd-order nonlinear coefficients, respectively \cite{Borisenko2018APL, Avci2018}.
The fitting gives parameters of 
$(a_{\rm USMR}, c_{\rm USMR}) = (1.08 \times 10^{-3}, 5.67 \times 10^{-6})$, 
resulting in the ratio 
$\xi_{\rm USMR} = c_{\rm USMR} / a_{\rm USMR} \sim 5.25 \times 10^{-3}$ $\rm (mA)^{-2}$
as summarized in Table \ref{table1}.
The fitting curve 
represented by a black solid line in Fig. \ref{usmr}(d) 
reproduces well the experimental results.
This indicates that the USMR in Fig. \ref{usmr}(d)
is traced back to the magnon generation/annihilation.

\subsection{Magnetization measurement}
Magnetization of the Pt-Py bilayer are measured using 
vibrating sample magnetometer (VSM).
Figure  \ref{mh}(a) shows 
magnetization curves   
at 296 (room temperature, black) and 346 K (red). 
The magnetization is normalized 
by the saturation magnetization $\mu_{0} M_{\rm S}$ at 296 K. 
By elevating the temperature from 296 K to 346 K, 
$\mu_{0} M_{\rm S}$ decreases slightly. 
In Fig. \ref{mh}(b), 
normalized $\mu_{0} M_{\rm S}$ 
is plotted as a function of temperature (red circles). 
$\mu_{0} M_{\rm S}$ 
decrease monotonically 
as temperature increases. 
The solid black line 
is a linear function 
obtained by the fitting of red circles. 
The gradient of the linear function 
is $-1.5\times10^{-3}$  $K^{-1}$. 
When sample temperature is 316 K, 
corresponding to 20 K elevation 
from room temperature, 
the normalized $\mu_{0} M_{\rm S}$ is 0.97.

\section{Discussion}
The 20 K temperature elevation 
due to the Joule heating 
confirmed in the USMR study  
causes 
a saturation magnetization decrease by 3 $\%$ 
evaluated by magnetization measurements.
This value is smaller than 
the decrease in $|\mu_{0}\Delta M_{\rm eff}|$ of 60 mT 
evaluated in Fig. \ref{anis}
corresponding to 9 $\%$ of $\mu_{0} M_{\rm{s}} =$ 658 mT.
Moreover, 
$|\mu_{0}\Delta M_{\rm eff}|$ variation 
is asymmetric for the sign of $I_{\rm dc}$.
Therefore,
the Joule heating is not dominant,
indicating that
the nonlinear spin polarization \cite{Xiao2022PRL, Xiao2023PRL}
causes the nonlinear spin torque 
observed in the present ST-FMR study.

The spin polarization 
${\delta}s_{i}$  
is described by 
an electronic response 
to an applied electric field $E_{j}$ as
${\delta}s_{i} = \chi_{ij}^{s (1)}E_{j} + \chi_{ijk}^{s (2)}E_{j}E_{k}$,
where $i$, $j$, and $k$ are Cartesian indices 
and the Einstein summation convention is adopted \cite{Xiao2022PRL, Xiao2023PRL}.
The $\chi_{ijk}^{s (2)}$ 
corresponds to the 2nd-order nonlinear response tensor,
which is relevant to the Berry connection polarizability
determined by the electronic band structures.
In Fig. \ref{result}, we focus on the nonlinear variation of $\Delta \alpha$ 
because $\Delta \alpha$ is relevant to the magnitude of spin torque.
The nonlinear spin polarization in Pt 
is partially absorbed by the Py magnetization, 
resulting in a nonlinear torque ($\propto \delta{\bm s} \times {\bf M}$). 
In this way,
the nonlinear torque  
caused by the nonlinear spin polarization
gives rise to  $\Delta \alpha$ depending on $(I_{\rm dc})^2$.
Indeed, 
the variation of 
$\Delta \alpha$ in Fig. \ref{result}
is reproduced well 
by the fitting curve 
$a_{\rm damp} I_{\rm dc} + b_{\rm damp} (I_{\rm dc})^2$
as represented by the black solid lines 
with $(a_{\rm damp}, b_{\rm damp}) 
= (1.35 \times 10^{-3}, 2.10 \times 10^{-5}) $.
An index of nonlinearity is evaluated to be
$\eta_{\rm damp} = b_{\rm damp} / a_{\rm damp} \sim 1.54 \times 10^{-2}$ $\rm (mA)^{-1}$
as shown in Table \ref{table1}. 

Based on the Holstein-Primakoff picture \cite{Holstein1940-vb},
the magnon number 
$\langle \hat{a} ^{\dag} \hat{a} \rangle$
is related to the precession angle $\varphi$ as
$\langle \hat{a} ^{\dag} \hat{a} \rangle 
= S (1-\cos \varphi) = 2S \sin(\varphi/2)$,
where $ \hat{a} ^{\dag} (\hat{a})$ 
is the creation (annihilation) operators for magnons
and the $S$ is the magnitude squared of the spin angular momentum \cite{Nakata2017-ps}.
Hence, the magnon generation (annihilation),
$\delta \langle \hat{a} ^{\dag} \hat{a} \rangle$,
by the spin injection 
gives rise to the increase (decrease) of the precession angle $\delta \varphi$,
which corresponds to
shrinkage (expansion) of the effective magnetization in the $x$-axis direction 
as in Figs. \ref{usmr}(a) and \ref{usmr}(b) \cite{Nakata2017-ps}.
Because the spin torque is expressed as
$\propto {\delta \bm s} \times {\bf M}_{\rm eff}$,
the nonlinear magnon generation/annihilation 
confirmed by the USMR study 
and followed by the magnetization shrinkage/expansion 
can thus be another origin of the nonlinear spin torque.

Given that the magnetization shrinkage/expansion 
contains both the 2nd- and 3rd-order nonlinearity \cite{Borisenko2018APL},
$\mu_{0}\Delta M_{\rm eff}$ 
is expressed as 
$ a_{\rm mag}  I_{\rm dc} + b_{\rm mag} (I_{\rm dc})^2 + c_{\rm mag} (I_{\rm dc})^3$,
where $a_{\rm mag}$, $b_{\rm mag}$ and $c_{\rm mag}$ 
correspond respectively 
to the linear coefficient, 
and the 2nd- and 3rd-order nonlinear coefficients. 
A fitting curve  
with $(a_{\rm mag}, b_{\rm mag}, c_{\rm mag}) 
= (9.51 \times 10^{-4}, 4.34 \times 10^{-5}, 2.35 \times 10^{-6}) $
represented by a solid black line in Fig. \ref{anis}
reproduces experimentally obtained values of $\mu_{0}\Delta M_{\rm eff}$. 
The $\xi_{\rm mag} = c_{\rm mag} / a_{\rm mag}$  
is $2.47 \times 10^{-3} $ ($\rm mA)^{-2}$, 
which is similar to 
$\xi_{\rm USMR} \sim  5.25 \times 10^{-3}$ ($\rm mA)^{-2}$
obtained from USMR measurements.
In addition,
the $\eta_{\rm mag} = b_{\rm mag} / a_{\rm mag}$
is $ \sim 4.56 \times 10^{-2}$ $\rm (mA)^{-1}$,
which is similar to 
$\eta_{\rm damp} \sim 1.54 \times 10^{-2}$ $\rm (mA)^{-1}$
as summarized in Table \ref{table1}.

\begin{table}[tb!]
\caption{\label{table1}
Nonlinearity indices $\eta$ and $\xi$ evaluated from ST-FMR and USMR measurements.
}
\begin{ruledtabular}
\begin{tabular}{lcc}
\textrm{ }&
\textrm{$\eta$ $(\rm mA)^{-1}$}&
\textrm{$\xi$ $(\rm mA)^{-2}$}\\
\colrule
USMR & - & $5.25 \times 10^{-3}$ \\
ST-FMR magnetization & $4.56 \times 10^{-2}$ & $2.47 \times 10^{-3}$ \\
ST-FMR damping & $1.54 \times 10^{-2}$ & - \\
\end{tabular}
\end{ruledtabular}
\end{table}

\begin{figure*}[tb!]
\includegraphics[width=16.0truecm,clip]{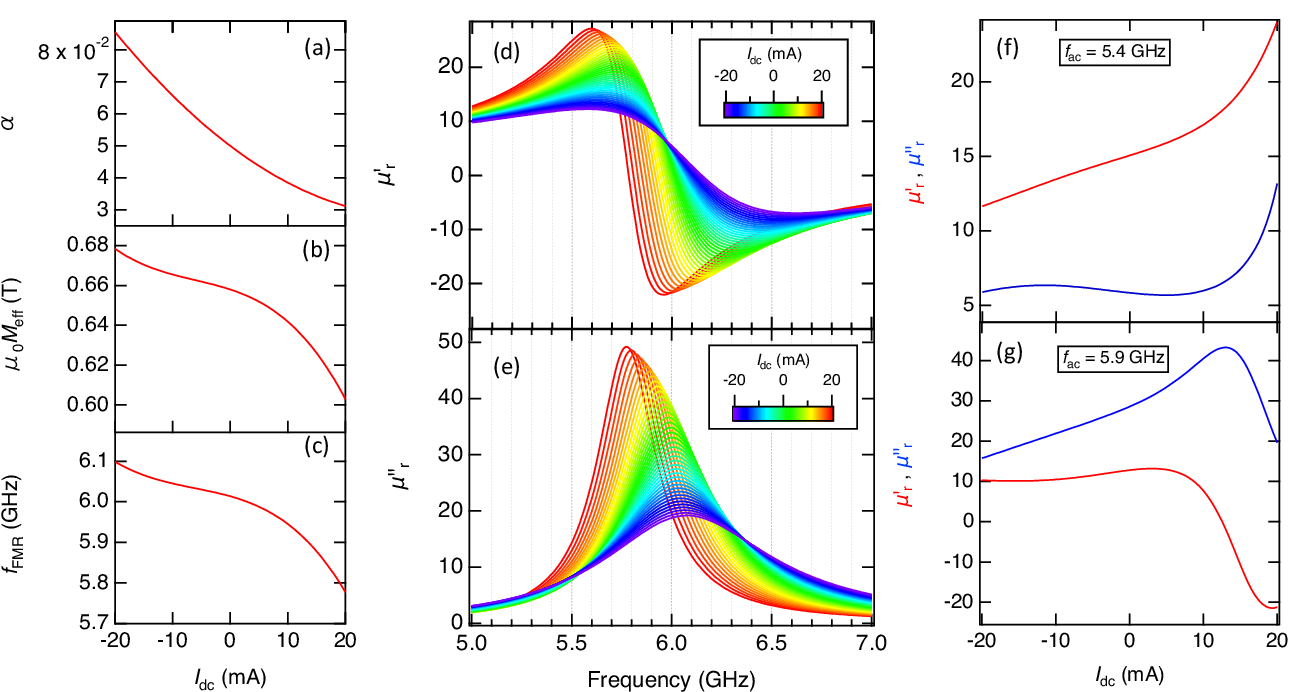}
\caption{
(a)-(c)
Damping parameter $\alpha$ calculated using nonlinear index of $\eta=1.54\times10^{-2}$, 
 effective magnetization $\mu_{0}M_{\rm eff}$, 
 and resonance frequency $f_{\rm FMR}$ 
 calculated using nonlinear index of $\eta=4.56\times10^{-2}$ and $\xi=2.47\times10^{-3}$ 
 are plotted as a function of $I_{\rm dc}$. 
 (d) Real $\mu_{\rm r}'$ and (e) imaginary $\mu_{\rm r}''$ parts of the relative magnetic permeability
 are evaluated using $\alpha$ and $\mu_{0}M_{\rm eff}$, 
 and plotted as a function of frequency. 
 Colors correspond to $I_{\rm dc}$ values. 
 The $\mu_{\rm r}'$ and $\mu_{\rm r}''$ versus $I_{\rm dc}$ 
 at (f) 5.4 GHz and (g) 5.9 GHz are also shown.
}
\label{Mu}
\end{figure*}

The present paper reveals that 
massive $I_{\rm dc}$ causes the nonlinear spin torque.
The nonlinear spin torque has two origins at least.
The first origin is electronic one, i.e., 
the nonlinear spin polarization in the Pt layer. 
The nonlinear spin polarization results in 
nonlinear spin torque 
directly observed by nonlinear $\Delta \alpha$ variation 
and represented by $\eta_{\rm damp}$ and $\eta_{\rm mag}$.
The second origin is magnonic one, i.e.,
the nonlinear magnon generation/annihilation 
confirmed by USMR and effective magnetization shrinkage/expansion, 
and represented by $\xi_{\rm USMR}$ and $\xi_{\rm mag}$.
In comparison between $\eta$ and $\xi$, 
Table \ref{table1} indicates that 
nonlinear spin polarization 
is dominant 
rather than nonlinear magnon generation/annihilation.
However, 
the nonlinear magnon generation/annihilation 
is caused by the linear spin polarization 
as well as nonlinear spin polarization \cite{Sandweg2011}. 
Therefore, 
the magnonic origin 
could be comparable to the electronic origin.
This is a future issue for the theoretical consideration.

The indices $\eta$ and $\xi$ 
obtained in ST-FMR and USMR measurements
are utilized in nonlinear spintronics of magnetic insulators \cite{chiba2014prappl}. 
The nonlinear spin polarization
is anticipated to bring about 
spin-torque oscillation in the Pt-Py bilayer \cite{Divinskiy2019, Fulara2020-ka}.
Furthermore, 
the nonlinear variation of 
the $\alpha$ and $\mu_{0}M_{\rm eff}$ 
due to the nonlinear spin torque
enables us to vary significantly 
the magnetic permeability 
of the magnetic bilayer system.
Figures \ref{Mu}(a) shows 
nonlinear variation of $\alpha$ 
calculated using
$\eta=1.54\times10^{-2}$. 
Nonlinear variation of $\mu_{0}M_{\rm eff}$
calculated using
$\eta=4.56\times10^{-2}$ and $\xi=2.47\times10^{-3}$ 
is shown in Fig. \ref{Mu}(b).
The relationship 
between $\mu_{0}M_{\rm eff}$ and the FMR resonance frequency $f_{\rm FMR}$ 
is expressed by the Kittel equation as
\begin{eqnarray}
\label{kittel2}
2\pi f_{\rm FMR}=\omega_{\rm FMR}=\gamma\sqrt{\mu_{0}H_{\rm ext}(\mu_{0}H_{\rm ext}+\mu_{0}M_{\rm eff})},
\end{eqnarray}
where $\omega_{\rm FMR}$ 
is the resonance angular frequency.
Nonlinear variation of $\omega_{\rm FMR}$ 
is thus obtained as plotted in Fig. \ref{Mu}(c).
Using $\alpha$, $\mu_{0} M_{\rm eff}$ and $\omega_{\rm FMR}$,
we evaluate
the magnetic permeability variation 
under a dc external magnetic field $\mu_{0}H_{\rm FMR}=58.4$ mT.
The relative permeability is written by 
real $\mu_{\rm r}'$ and 
imaginary  $\mu_{\rm r}''$ parts \cite{Kodama2023} as
\begin{eqnarray}
\label{relativemu}
\mu_{\rm r}(\omega)=\mu_{\rm r}'(\omega)-j\mu_{\rm r}''(\omega), 
\end{eqnarray}
where
\begin{subequations}
\label{mur}
\begin{align}
\mu_{\rm r}' (\omega) &=1 +  \gamma\mu_{0} M_{\rm eff} \frac{\omega_{\rm FMR} (\omega_{\rm FMR}^{2} - \omega^{2}) + \omega_{\rm FMR} \omega^{2}\alpha^{2}}
{[\omega_{\rm FMR}^{2} - \omega^{2} (1 + \alpha^{2})]^{2} + 4 \omega_{\rm FMR}^{2} \omega^{2} \alpha^{2}} , \\
\mu_{\rm r}'' (\omega) &= \gamma\mu_{0} M_{\rm eff} \frac{\alpha \omega [\omega_{\rm FMR}^{2} - \omega^{2} (1 + \alpha^{2})]}
{[\omega_{\rm FMR}^{2} - \omega^{2} (1 + \alpha^{2})]^{2} + 4 \omega_{\rm FMR}^{2} \omega^{2} \alpha^{2}}.
\end{align}
\end{subequations}
By substituting $\alpha$, $\mu_{0} M_{\rm eff}$ and $\omega_{\rm FMR}$ 
into the Eq. (\ref{mur}),
$\mu_{\rm r}(\omega)$ at each $I_{\rm dc}$ 
is evaluated. 

Figures \ref{Mu}(d) and \ref{Mu}(e)
show dispersion curves of $\mu_{\rm r}'(\omega)$ and  $\mu_{\rm r}''(\omega)$, 
respectively, at various $I_{\rm dc}$ from $-$20 (blue) to $+$20 mA (red)
at 1 mA intervals.
In Fig. \ref{Mu}(f), 
$\mu_{\rm r}'$ and $\mu_{\rm r}''$ obtained at 5.4 GHz
are plotted as a function of $I_{\rm dc}$.
When $I_{\rm dc}$ is varied from $-$20 to 10 mA,
only the $\mu_{\rm r}'$ can be modified.
Contrastingly, 
at 5.9 GHz, 
only $\mu_{\rm r}''$ can be modified as in Fig. \ref{Mu}(g). 
This is an advantageous 
in realizing time-varying permeability metamaterials
for the microwave frequency conversion 
towards 6th-generation mobile communication light source.

\section{Conclusion}
We directly observe 
the nonlinear spin torque in the Pt-Py bilayer 
by means of ST-FMR with a large dc current.
The nonlinear spin torque 
observed by nonlinear $\Delta \alpha$ variation 
is attributed primarily to 
nonlinear spin polarization 
represented by $\eta$.
Moreover, 
ST-FMR and USMR measurements 
demonstrate that 
nonlinear magnon generation/annihilation 
followed by shrinkage/expansion of effective magnetization, 
represented by $\xi$,
is another origin of the nonlinear spin torque. 
Comparison between $\eta$ and $\xi$ 
indicates that 
nonlinear spin polarization 
is dominant 
rather than nonlinear magnon generation/annihilation.
The real and imaginary parts of permeability 
can be varied independently 
using nonlinear spin torque.
The present paper 
paves a way to 
spin-Hall effect based nonlinear spintronic devices 
as well as time-varying nonlinear magnetic metamaterials 
with tailor-made permeability.

\section*{Acknowledgements}
The authors acknowledge 
Masatoshi Hatayama, 
Takashi Komine,
Saburo Takahashi
and Yoshiaki Kanamori 
for their valuable contributions in this work.
N. K., S. O., and S. T. also thank 
Network Joint Research Center for Materials and Devices (NJRC).
This work is financially supported by JST- CREST (JPMJCR2102).

%\clearpage

%\bibliographystyle{apsrev4-2}
%\bibliographystyle{planenat}
%\bibliography{refprappl}% Produces the bibliography via BibTeX.
%\bibliographystyle{plane}

\end{document}